\documentstyle[12pt]{article}
\begin{document}
{\hfill PUTP-94-26}

\vspace{8mm}
\centerline{\large\bf $\xi(2230)$ IS LIKELY TO BE A GLUEBALL
\footnote{Talk presented at the CCAST Workshop on Tau-Charm Factory,
Beijing, November 1994} }

\vspace{10mm}
\centerline{Kuang-Ta Chao}
\vspace{4mm}
\centerline{\it CCAST (World Laboratory), ~Beijing 100080,~P.R.China}
\centerline{\it and}
\centerline{\it Department of Physics, ~Peking University,
         ~Beijing 100871,~P.R.China}

\vspace{8mm}
\begin{abstract}
On the basis of the recent results of $\xi(2230)\rightarrow \pi^+\pi^-,
 ~p\bar p$ and $\xi(2230)\rightarrow K^+K^-,~K_SK_S$, measured
by the BES Collaboration in $J/\psi$ radiative decays,
combined with the PS185 experiment of $p\bar p\rightarrow
\xi(2230)\rightarrow K\bar K$, we argue that because of
its very narrow partial decay widths to $\pi\pi$ and $K\bar K$
( less than $1~MeV$ ),
its large production rate in $J/\psi$ radiative decays,
and its flavor-symmetric couplings to $\pi\pi$ and $K\bar K$,
the $\xi(2230)$ is very likely to be
a $J^{PC}=(even)^{++}$ glueball.
\end{abstract}
\vspace{6mm}

\newpage

The $\xi(2230)$ particle, a narrow state with a width of about 20 MeV,
was first observed by the MARK III Collaboration
at SLAC-SPEAR in the radiative decays of $J/\psi$ to $K^+K^-$ and
$K_SK_S$ final states$^{[1]}$.
There are many speculations about the nature of $\xi(2230)$.
It could be an ordinary $s\bar s$ quarkonium state$^{[2]}$,
a baryonium
state such as $(qs)_{3^*}-({\bar q}{\bar s})_3$ with $q$ being a $u$
or $d$ quark$^{[3]}$, a $\Lambda{\bar \Lambda}$ bound state$^{[4]}$,
an $ss\bar s\bar s$ state$^{[5]}$, a ${\bar q}qg$ hybrid$^{[6,3]}$, or
a glueball$^{[7]}$.

Recently, the BES Collaboration at BEPC (Beijing Electron Positron
Collider) has reported new results of the $\xi(2230)$. Two significant
decay modes $\xi\rightarrow\pi^+\pi^-$ and $\xi\rightarrow p\bar p$ have been
observed in addition to $\xi\rightarrow K^+K^-,~K_SK_S$. The measured mass,
width, and branching ratios are as follows$^{[8]}$.

In $J/\psi\rightarrow \gamma\xi,~\xi\rightarrow \pi^+\pi^-$:
\begin{eqnarray}
&&M_{\xi}=(2235\pm 4\pm 6)MeV, \nonumber\\
&&\Gamma_{\xi}=(19^{+13}_{-11}\pm 12)MeV, \nonumber\\
&&BR(J/\psi\rightarrow \gamma\xi)\times BR(\xi\rightarrow \pi^+\pi^-)=
(5.6^{+1.8}_{-1.6}\pm 1.4)\times 10^{-5}.
\end{eqnarray}

In $J/\psi\rightarrow \gamma\xi,~\xi\rightarrow p\bar p$:
\begin{eqnarray}
&&M_{\xi}=(2235\pm 4\pm 5)MeV, \nonumber\\
&&\Gamma_{\xi}=(15^{+12}_{-9}\pm 9)MeV, \nonumber\\
&&BR(J/\psi\rightarrow \gamma\xi)\times BR(\xi\rightarrow p\bar p)=
(1.5^{+0.6}_{-0.5}\pm 0.5)\times 10^{-5}.
\end{eqnarray}

In $J/\psi\rightarrow \gamma\xi,~\xi\rightarrow K^+K^-$:
\begin{eqnarray}
&&M_{\xi}=(2230^{+6}_{-7}\pm 12)MeV, \nonumber\\
&&\Gamma_{\xi}=(20^{+20}_{-15}\pm 12)MeV, \nonumber\\
&&BR(J/\psi\rightarrow \gamma\xi)\times BR(\xi\rightarrow K^+K^-)=
(3.3^{+1.6}_{-1.3}\pm 1.1)\times 10^{-5}.
\end{eqnarray}

In $J/\psi\rightarrow \gamma\xi,~\xi\rightarrow K_SK_S$:
\begin{eqnarray}
&&M_{\xi}=(2232^{+8}_{-7}\pm 15)MeV, \nonumber\\
&&\Gamma_{\xi}=(20^{+25}_{-16}\pm 10)MeV, \nonumber\\
&&BR(J/\psi\rightarrow \gamma\xi)\times BR(\xi\rightarrow K_SK_S)=
(2.7^{+1.1}_{-0.9}\pm 1.0)\times 10^{-5}.
\end{eqnarray}

These results indicate that the $\xi(2230)$ decays are probably
flavor symmetric, and its total width is as narrow as about 20 MeV.
These features may make
$\xi(2230)$ distinguishable from many other states.

The situation will become most clear-cut when combining the BES result with
the experiments E789$^{[9]}$ at BNL-AGS,~PS170$^{[10]}$ at CERN-LEAR,
and PS185$^{[11]}$ at CERN-LEAR aimed at $\xi(2230)$ in
the reaction $p\bar p\rightarrow\xi(2230)\rightarrow K\bar K$.
These experiments find no any significant structure in the $\xi(2230)$
region and only give an upper bound for the following product
of branching ratios (see, e.g., refs.[11,12])
\begin{equation}
BR(\xi\rightarrow p\bar p)\times BR(\xi\rightarrow K\bar K)
<1.5\times 10^{-4},
\end{equation}
where $J=2, ~\Gamma_{\xi}=20MeV$ for $\xi(2230)$ are assumed,
and where $K\bar K$ include all kaon pairs.

Note that $BR(\xi\rightarrow p\bar p)$ measured by BES is only
four times smaller than $BR(\xi\rightarrow K\bar K)$. Then
combining the BES branching ratios given in (1)-(4) with
the PS185 result (5), and
taking again $\Gamma_{\xi}=20MeV$, we find
\begin{eqnarray}
&&BR(\xi\rightarrow\pi^+\pi^-)<2.4\times 10^{-2},~~~~~\Gamma(\xi
\rightarrow\pi^+\pi^-)<480KeV,\\
&&BR(\xi\rightarrow p\bar p)<0.6\times 10^{-2},~~~~~~~~~\Gamma(\xi
\rightarrow p\bar p)<120KeV,\\
&&BR(\xi\rightarrow K^+K^-)<1.4\times 10^{-2},~~~\Gamma(\xi
\rightarrow K^+K^-)<280KeV,\\
&&BR(\xi\rightarrow K_SK_S)<1.1\times 10^{-2},~~~~\Gamma(\xi
\rightarrow K_SK_S)<220KeV.
\end{eqnarray}
In particular, with (1) and (6), we immediately get a rather large
branching ratio for $J/\psi \rightarrow \gamma\xi(2230)$
\begin{equation}
BR(J/\psi\rightarrow \gamma\xi)>2.3\times 10^{-3}.
\end{equation}

With above observed bounds, we can see at least three
pronounced features of the $\xi$.

(1) The partial decay widths of $\xi$ to $\pi^+\pi^-$ and $K\bar K$,
as shown in (6)-(9), are much smaller
than that of any conventional $q\bar q$ states,
including $(u\bar u+d\bar d)$, and $s \bar s$, or their admixtures.
It is obvious that for any conventional $q\bar q$ meson whose
decay is OZI-rule allowed, its total width and partial widths of
certain main decay modes must be of order of $10-100MeV$.
E.g., for the $P$-wave $J^{PC}=2^{++}$ mesons, the $f_2(1270)$,
which is mainly a $(u\bar u+d\bar d)$ state, has
a patial decay width  of about $150MeV$ to $\pi\pi$, while the
$f_2'(1525)$, which is mainly an $s\bar s$ state, has a partial
decay width  of about $50MeV$ to $K\bar K$$^{[12]}$.
For the $F$-wave mesons, based on some quark model calculation
it was argued$^{[2]}$
that if $\xi(2230)$ was a $^3F_2$ or $^3F_4$
$s\bar s$ state, its decay to $K\bar K$ could be suppressed
by the $L=3$ centrifugal barrier and consequently the decay width
to $K\bar K$ could be lowered to $20-30MeV$, but can not be as
small as , say, $500KeV$. On the other hand,
the $f_4(2050)$, which is a $(u\bar u+d\bar d)$ dominated
$^3F_4$ state, has an observed total width of $200MeV$ and a partial
decay width of $30MeV$ to $\pi\pi$ $^{[12]}$. We see that, with both
experimental observations and quark model calculations,
all this kind of $q\bar q$ states
can hardly have a total width of $20 MeV$ and, in particular,
can not have a partial decay width of, say,
$500KeV$ to $\pi\pi$ or $K \bar K$, as observed for $\xi(2230)$.
Therefore, as the result of observed small partial widths of $\xi(2230)
\rightarrow\pi\pi,~K\bar K$,  we may
conclude that the $\xi(2230)$ can not be a conventional $q\bar q$
meson.

(2) The $\xi(2230)$ has a large production rate in $J/\psi$
radiative decays which are known as the gluon-rich channels.
In fact,
according to (10) and the data of $J/\psi$ radiative
decays$^{[12]}$, the production rate of $\xi(2230)$
could be only smaller than $\eta_c$ and $\eta'(958)$,
but larger than or as large as $\iota(1440)$, $\theta(1710)$,
$f_4(2050)$, and $f_2(1270)$. As discussed
above, due to the small partial widths of $\xi\rightarrow
\pi\pi,~K\bar K$, $\xi$ can hardly be the counterpart of $f_4(2050)$ or
$f_2(1270)$, or other $q\bar q$ mesons. As for the $q\bar qq\bar q$
states or $\Lambda\bar \Lambda$-like baryon-antibaryon bound states,
according to the naive quark pair counting rule, they are
usually expected
to have smaller production rates than the corresponding $q\bar q$
states, since the creation of
more quark pairs is needed for $q\bar qq\bar q$ or baryon-antibaryon
bound state production.
Most naturally, the rich production of $\xi$ in $J/\psi$ radiative
decays will imply that the $\xi(2230)$ is likely to be
a glueball or a $q\bar qg$ hybrid state, but the former should
have an even larger production rate than the latter.

(3) The $\xi(2230)$ decays are probably flavor-symmetric with
many decay modes. The closeness of observed decay branching
ratios of $\xi$ to $\pi\pi$
and $K\bar K$, as shown in (1)-(4), apparently suggests a possible
flavor-singlet nature of the $\xi$, while the smallness of
these branching ratios, as shown in (6)-(9), indicates that $\xi$
may have as many as tens decay modes. These two features bear
resemblance to the charmonium decays, in particular, to the $\chi_{c0}$
and $\chi_{c2}$ decays. Both $\chi_{c0}$ and $\chi_{c2}$ decays
may proceed via two steps: first the $c\bar c$ pair annihilate into two
gluons; and then the two gluons are hadronized into light mesons and baryons.
The gluon hadronization is flavor-symmetric and then leads to
flavor-symmetric decays. This picture is stronly supported
by the $\chi_{c0}$ and $\chi_{c2}$ decays. E.g., the $\chi_{c2}$
is found to have approximately the same decay rate to $\pi^+\pi^-$
as to $K^+K^-$, and the same decay rate to $\pi^+\pi^-\pi^+\pi^-$
as to $\pi^+\pi^-K^+K^-$$^{[12]}$. For a glueball, say, a
$J^{PC}=2^{++}$ glueball, which is made of two gluons, its decay
proceeds via the two-gluon hadronization, which is similar to the
second step of the $\chi_{c2}$ decay. The difference between the
glueball and $\chi_{c2}$ in their decays is that the two gluons
are hadronized at different energy scales, and consequently in the
two cases the branching ratio for a given final state can be different.
At the higher energy scale like the $\chi_{c2}$ mass, more channels
are open and competing, and more particles (pions mainly) are
produced with certain averaged momenta to balance the primitive
leading particles converted by the gluons, and therefore
the decay branching ratio to $\pi^+\pi^-\pi^+\pi^-$ can be larger
than to $\pi^+\pi^-$, and that to $p\bar p\pi^+\pi^-$ can be
larger than $p \bar p$, as observed experimentally in the
$\chi_{c2}$ decays$^{[12]}$. Despite of this difference between
the charmonium and glueball,
we believe
the observed flavor-symmetric pattern of charmonium decays does lend
strong support to the conjecture that the glueball decays should
be flavor-symmetric.
Another possible feature of the glueball decay is that glueballs probably
have more decay modes than conventional $q\bar q$ states. A $q\bar q$
meson decay occurs when the color flux tube formed by $q$ and $\bar q$
is broken at large distances
by the creation of new quark pairs (the OZI allowed decays);
whereas a glueball decay proceeds via the gluon hadronization.
There are more possibilities and combinations for the gluon
fragmentation and hadronization
than for the quark pair creation in a color flux tube.
Therefore, a glueball may have more decay modes than a $q\bar q$
meson, and hence have smaller branching ratios to many final states.
In these connections, for $\xi(2230)$
the observed flavor-symmetric decays to
$\pi\pi,~K\bar K$ and the smallness of these decay branching ratios
seem to favor the assignment that the $\xi(2230)$ is a glueball.
A $(u\bar u+d\bar d)g$ habrid state may also have comparable strengths
to couple to
$\pi\pi$ and $K\bar K$, but it can be distinguished from a glueball
by certain special decay modes, e.g., it can decay to $\omega\phi$ but
not to $\phi\phi$.
The flavor-symmetric couplings of $\xi$ to $\pi\pi$ and $K\bar K$
also disfavor the $q\bar q$ states, because flavor mixings should be
small for orbitally excited $q\bar q$ mesons. E.g., $f_2(1270)$
and $f_2'(1525)$ $(L=1)$ have dominant decay modes to $\pi\pi$ and
$K\bar K$ respectively; and the $f_4(2050)$ $(L=3)$ has dominant
decay modes to $\omega\omega,~\pi\pi$, while its branching ratio
to $K\bar K$ is only $7\times 10^{-3}$$^{[12]}$.

With the three observations made above, we see that the $\xi(2230)$
is very unlikely to be a conventional $q\bar q$ meson,
less likely to be a four-quark state or a baryon-antibaryon bound state,
but very likely to be a $J^{PC}=(even)^{++}$ glueball, though a $q\bar qg$
habrid could also be a disfavorable possibility.
To draw a more definite conclusion about the
nature of $\xi(2230)$, further experimental studies should be done.
Following suggestions might be useful.

Searching for more decay modes of $\xi(2230)$.
Since the observed $\pi\pi,
K\bar K$, and $p\bar p$ are expected to be, according to (6)-(9),
only a small portion of the decay modes of $\xi$, other decay modes
such as $\eta\eta, ~\eta\eta',~\eta'\eta'$ and
$\pi\pi\pi\pi,~\pi\pi K\bar K, ~\rho\rho,~K^*{\bar K^*}, ~\phi\phi$
may be important. A systematical test of the flavor
symmetric nature in the decays will be crucial for the glueball
interpretation of $\xi$.

Searching for some special decay modes of $\xi(2230)$, e.g., $\omega
\phi$ and $\phi\phi$. A glueball can decay to $\phi\phi$ but not
$\omega\phi$, whereas a $(u\bar u+d\bar d)g$ habrid may decay to
$\omega\phi$ but not $\phi\phi$. Therefore those modes may provide
crucial tests for distinguishing between the glueball and the habrid.
The $\omega\phi$ mode is also a prediction for the $qs\bar q\bar s$
state.

Determining the spin-parity of $\xi(2230)$. The spin-parity of $\xi$
could be $0^+,2^+$ or $4^+$.
$J^P=4^+$ will disfavor the glueball and habrid
interpretations, because it requires a non-$S$ wave orbital angular
momentum between the constituents, and then lead to higher mass
and lower production rate in $J/\psi$ radiative decays than $\xi(2230)$.
In this connection, it is interesting to note that a comprehensive
lattice study of SU(3) glueballs by the UKQCD Collaboration suggests
the mass of the $2^{++}$ glueball be $2270\pm 100MeV$$^{[13]}$.
Therefore, if the spin of $\xi$ is experimentally determined to be 2,
then the
$2^{++}$ gluball interpretation fo $\xi$ will be even more strongly
supported.

Examining the inclusive photon spectrum in the $\xi(2230)$ region
in the $J/\psi$ radiative
decays. According to (5), the $\xi$ has a large production
rate in $J/\psi\rightarrow \gamma gg\rightarrow\gamma+hadrons$
and therefore
should show up as a narrow peak in the inclusive photon spectrum.
This will be a test of consistency for the BES$^{[8]}$ and
PS185$^{[11]}$ experiments, and is also very important for the
understanding of the nature of $\xi(2230)$.

To sum up, we believe that the recent result reported by BES
is very encouraging in the idenfication of the puzzling state
$\xi(2230)$. With both the BES and PS185 experiments, this particle
is found to have striking features that
it has very narrow partial decay widths to
$\pi\pi$ and $K\bar K$, a large production rate in $J/\psi$
radiative decays, and  flavor-symmetric couplings to $\pi\pi$
and $K\bar K$.
These features strongly favor the conclusion that the $\xi(2230)$
is likely to be a $J^{PC}=(even)^{++}$ glueball.

\bigskip
I would like to thank T.Huang, M.H.Ye, and Z.P.Zheng, the organizers
of the CCAST Workshop on Tau-Charm Factory, Beijing, November 1994,
where this work was presented. I also thank J.Li, J.H.Liu, W.G.Yan,
S.Z.Ye, Z.P.Zheng, Y.C.Zhu, and, in particular, S.Jin for many stimulating
discussions on the experimental status of $\xi(2230)$.
This work was supported in part by the National Natural
Science Foundation of China, and the State Education
Commission of China.

\end{document}